\documentclass[aps,prb,reprint,showpacs]{revtex4-2}
\usepackage[dvipdfmx]{graphicx}
\usepackage{subfigmat}
\usepackage{amsmath,amssymb}
\usepackage{color}
\usepackage{bm}
\usepackage{hyperref}
\usepackage{comment}
\usepackage[normalem]{ulem}

\usepackage{cancel}

\usepackage{upgreek}

\begin{document}

\title{Spin Pumping into Carbon Nanotubes}

\author{K. Fukuzawa$^{1}$, T. Kato$^{1}$, M. Matsuo$^{2,3,4,5}$, T. Jonckheere$^{6}$, J. Rech$^{6}$, and T. Martin$^{6}$}
\affiliation{
${^1}$Institute for Solid State Physics, The University of Tokyo, Kashiwa, 277-8581, Japan\\
${^2}$Kavli Institute for Theoretical Sciences, University of Chinese Academy of Sciences, Beijing, 100190, China\\
${^3}$CAS Center for Excellence in Topological Quantum Computation, University of Chinese Academy of Sciences, Beijing, 100190, China\\
${^4}$Advanced Science Research Center, Japan Atomic Energy Agency, Tokai, 319-1195, Japan\\
${^5}$RIKEN Center for Emergent Matter Science (CEMS), Wako, Saitama, 351-0198, Japan\\
${^6}$Aix Marseille Univ, Universit\'e de Toulon, CNRS, CPT, IPhU, AMUtech, Marseille, France
}

\date{\today}

\begin{abstract}
We theoretically study spin pumping from a ferromagnetic insulator (FI) into a carbon nanotube (CNT) .
By employing the bosonization method, we formulate the Gilbert damping induced by the FI/CNT junction, which can be measured by ferromagnetic resonance.
We show that the increase in the Gilbert damping has a temperature dependence characteristic of a Luttinger liquid and is highly sensitive to the Luttinger parameter of the spin sector for a clean interface.
We also discuss the experimental relevance of our findings based on numerical estimates, using realistic parameters.
\end{abstract}
\maketitle 

\section{Introduction}
\label{sec:introduction}

Spin pumping induced by ferromagnetic resonance (FMR)~\cite{Tserkovnyak2002,Hellman2017} is a fundamental technique in spintronics for generating spin current from a ferromagnet to an adjacent material~\cite{Zutic2004,Tsymbal2019}.
While spin pumping has been used for injecting spin into various materials, it can also be utilized for detecting spin excitations in various systems~\cite{Han2020,Yang2018,Qiu2016,Yamamoto2021,Ominato2020a,Ominato2020b,Yama2021,Inoue2017,Silaev2020,Silaev2020b,Ominato2022a,Ominato2022b,Funato2022}.
Compared with bulk measurement techniques, such as nuclear magnetic resonance (NMR) and neutron scattering experiments, spin pumping has an advantage in sensitivity for nanostructured systems such as surfaces, thin films and atomic-layer compounds~\cite{Han2020}.

The study of exotic spin excitations which emerge in specific materials is one of the forefront topics of condensed matter physics.
A typical example is spin excitation in quasi-one-dimensional interacting electron systems, whose low-energy excitation can be described by the Tomonaga-Luttinger liquid~\cite{Voit1995,vonDelft1998,Giamarchi2003}.
Spin excitations inherent to the Tomonaga-Luttinger liquid have been studied in carbon nanotubes (CNTs) by using NMR~\cite{Singer2005,Dora2007,Ihara2010}.
While NMR can detect the local spin susceptibility in CNTs, the use of spin pumping to detect spin excitations is expected to provide useful information  reflecting the exotic character of the Luttinger liquid, which cannot be captured by NMR.
It is thus important to clarify what kind of information about the Luttinger liquid can be obtained from a spin pumping experiment.

\begin{figure}[tbp]
  \begin{center}
    \includegraphics[clip, width=5.5cm]{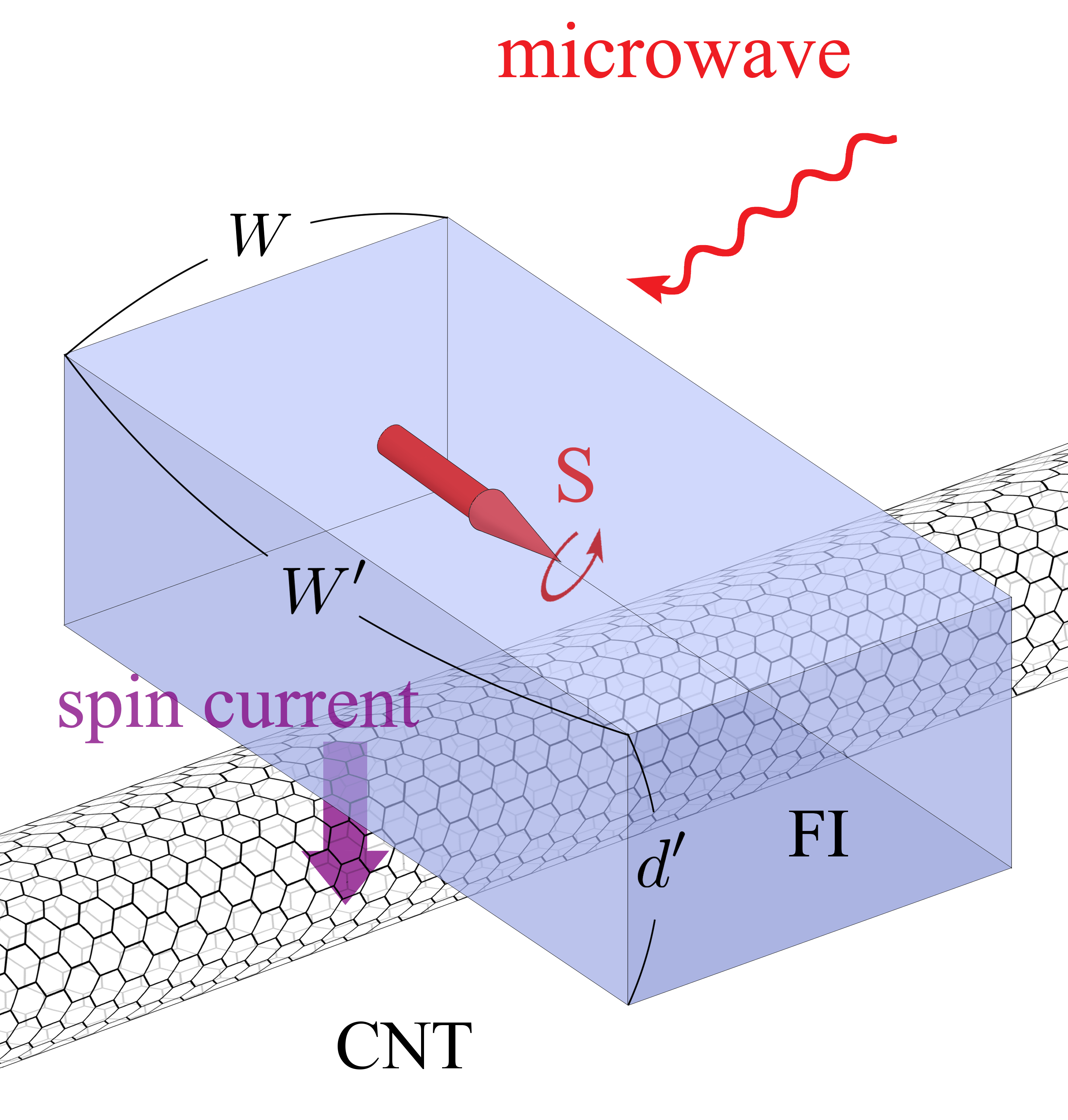}
    \caption{Magnetic junction composed of a ferromagnetic insulator (FI) and a single-wall carbon nanotube (CNT).
    The dimension of the FI is $W\times W'\times d'$.}
    \label{fig:spin-pumping-setup}
  \end{center}
\end{figure}

In this work, we theoretically formulate the increase in the Gilbert damping due to spin pumping in a setup in which spin is injected into CNTs. We consider a magnetic junction composed of a ferromagnetic insulator (FI) and a single-wall CNT (see Fig.~\ref{fig:spin-pumping-setup}) and take interfacial randomness into account with a simple model. We derive an analytic expression for the increase in the Gilbert damping by utilizing the bosonization method and second-order perturbation with respect to the interfacial exchange coupling. 

We will focus on the two limiting cases, i.e., a clean interface and a dirty interface.
We show that for both cases the temperature dependence of the increase of the Gilbert damping shows a power-law behavior, with an exponent reflecting the Luttinger parameters.
For a clean interface, the exponent includes information on the Luttinger parameters in the spin sector and is shown to be sensitive to small deviations from unity (which is the value of the SU(2) symmetric model in the spin sector).
For a dirty interface, the exponent depends on the Luttinger parameters of both the spin and charge sectors as in an NMR measurement.
We estimate the increase of the Gilbert damping using realistic parameters and discuss the experimental feasibility.

Our paper is organized as follows.
We introduce the microscopic model of the FI/CNT magnetic junction in Sec.~\ref{sec:Model}.
We analytically calculate the increase in the Gilbert damping in Sec.~\ref{sec:formuation} and subsequently estimate it with realistic parameters in Sec.~\ref{sec:result}.
Finally, we briefly discuss the experimental relevance of our findings in Sec.~\ref{sec:Relevance} and summarize our results in Sec.~\ref{sec:summary}.
A detailed derivation of the analytic expressions is given in the two Appendices.

\section{Model}
\label{sec:Model}

Let us consider a junction composed of a CNT and FI, whose Hamiltonian is given by $\mathcal{H} = \mathcal{H}_{\mathrm{CNT}} + \mathcal{H}_{\mathrm{FI}} + \mathcal{H}_{\mathrm{int}}$.
Here, $\mathcal{H}_{\mathrm{CNT}}$ and $\mathcal{H}_{\mathrm{FI}}$ describe electrons in the CNT and FI, respectively, and $\mathcal{H}_{\mathrm{int}}$ represents the interfacial exchange interaction between the CNT and FI.
We will give their explicit forms in the subsections that follow.

\subsection{Carbon nanotube}
\label{sec:CNTModel}

The low-energy Hamiltonian of electrons in CNTs is given by
\begin{align}
\mathcal{H}_{\mathrm{CNT}} &= \mathcal{H}_{\mathrm{K}} + \mathcal{H}_{\mathrm{C}} ,
\end{align}
where $\mathcal{H}_{\mathrm{K}}$ and $\mathcal{H}_{\mathrm{C}}$ represent the kinetic energy and the forward scattering potential due to the screened Coulomb interaction, respectively.
Using standard conventions~\cite{Egger1998}, the Hamiltonians describing these energies of electrons in CNTs are given by
\begin{align}
\mathcal{H}_{\mathrm{K}} &= -iv_{\rm F} \int dx \sum_{r\alpha\sigma} r\psi^+_{r\alpha\sigma}(x)\partial_x \psi_{r\alpha\sigma}(x) , \\
\mathcal{H}_{\mathrm{C}} & =\frac{1}{2}\int dx\,dy\,\rho(x) V(x-y) \rho(y) ,
\end{align}
where $\psi_{r\alpha\sigma}(x)$ is the slowly varying part of the field operator of electrons, $v_{\rm F}$ is the Fermi velocity, $V(x)$ is the screened Coulomb potential, and $\rho(x) = \sum_{r\alpha\sigma} \psi^\dagger_{r\alpha\sigma}(x) \psi_{r\alpha\sigma}(x)$ is the electron density operator.
The subscripts, $r$ ($=\pm$), $\alpha$ ($=\pm$), and $\sigma$ ($=\pm$), represent the direction of propagation, the nanotube branch (the valley), and the spin orientation, respectively.
Using the bosonization method~\cite{vonDelft1998,Egger1998}, the annihilation operator describing fermions in the CNT can be expressed in terms of bosonic fields, $\theta_{\alpha\sigma}(x)$ and  $\phi_{\alpha\sigma}(x)$, as
\begin{align}
\psi_{r\alpha\sigma}(x)&=\frac{\eta_{r\alpha\sigma}}{\sqrt{2\pi a}}e^{i(-r\theta_{\alpha\sigma}(x)+\phi_{\alpha\sigma}(x))},
\end{align}
where $\eta_{r\alpha\sigma}$ is the Klein factor, and $a$ is a short-length cutoff which can be identified with the lattice constant of the CNT.
To diagonalize the Hamiltonian, we introduce new bosonic fields for the charge and spin sectors, $\theta_{j \delta}(x)$ and $\phi_{j \delta}(x)$ as
\begin{align}
\theta_{\alpha\sigma}(x)&=\frac{1}{2} \sum_{j\delta} h_{j\delta}(\alpha,\sigma) \theta_{j\delta}(x),\\
\phi_{\alpha\sigma}(x)&=\frac{1}{2} \sum_{j\delta} h_{j\delta}(\alpha,\sigma) \phi_{j \delta}(x),
\end{align}
where $\delta$ ($=\pm$) represents symmetric/antisymmetric modes, $j$ ($=c,s$) indicates the charge/spin mode, $h_{c+}=1$, $h_{c-}=\alpha$, $h_{s+}=\sigma$, and $h_{s-}=\alpha\sigma$.
The Hamiltonian of the CNTs can be written as
\begin{align}
\mathcal{H}_{\rm CNT}=\sum_{j,\delta}\frac{v_{j\delta}}{2\pi}\int \! dx[K_{j\delta}^{-1}(\partial_x\theta_{j\delta})^2+K_{j\delta}(\partial_x\phi_{j\delta})^2] ,
\end{align}
where $K_{j\delta}$ is the Luttinger parameter and $v_{j\delta}=v_F/K_{j\delta}$.

\subsection{Ferromagnetic insulator}
\label{sec:FImodel}

We consider a bulk FI described by the quantum Heisenberg model and employ the spin-wave approximation assuming that the temperature is much lower than the magnetic transition temperature and the magnitude of the localized spin, $S_0$, is much larger than one~\cite{Kato2019,Ominato2020a,Yamamoto2021,Yama2021,Ominato2022a,Ominato2022b,Funato2022}.
In this situation, the Hamiltonian for the FI is approximately written as a superposition of magnon modes:
\begin{align}
\mathcal{H}_{\rm FI}=\sum_{\bm{k}}\hbar \omega_{\bm{k}} b^\dagger_{\bm{k}} b_{\bm{k}}, 
\end{align}
where $b_{\bm{k}}$ is the annihilation operator of magnons, $\hbar\omega_{\bm{k}}=\mathcal{D} \bm{k}^2+\hbar\gamma_{g} h_{dc}$ is the magnon dispersion, $\mathcal{D}$ is spin stiffness, $\gamma_{g}$ is the gyromagnetic ratio, and $h_{dc}$ is the static magnetic field.
We will only focus on uniform spin precession induced by external microwaves.
For this purpose, it is sufficient to consider the magnon mode of ${\bm k}={\bm 0}$ with the simplified Hamiltonian
\begin{align}
\mathcal{H}_{\rm FI}=\hbar \omega_{\bm{0}} b^\dagger_{\bm{0}} b_{\bm{0}}.
\end{align}
Microwave absorption in FMR can be related to the imaginary part of the retarded spin correlation function, which is defined as
\begin{align}
G^R(\omega)=-\frac{i}{\hbar} \int^\infty_0 dt \, e^{i(\omega +i\delta)t}\langle[S_{\bm{0}}^+(t), S^-_{\bm{0}}]\rangle ,
\end{align}
where $S_{\bm{0}}^+ = \sqrt{2S_0} b_{\bm 0}$ and $S_{\bm{0}}^- = \sqrt{2S_0} b^\dagger_{\bm 0}$ are spin ladder operators of the FI for ${\bm k}={\bm 0}$ and $S_{\bm{0}}^+(t)=e^{i{\cal H}t/\hbar}S_{\bm{0}}^+e^{-i{\cal H}t/\hbar}$.
For an isolated bulk FI, the spin susceptibility is calculated as:
\begin{align}
G^R_0(\omega)=\frac{2S_0/\hbar}{\omega-\omega_{\bm{0}}+i\delta} .
\end{align}
In real experiments, the FMR linewidth is finite due to the the Gilbert damping. 
To represent this finite spin relaxation in the bulk FI, we introduce a phenomenological dimensionless parameter $\alpha_{\rm G}$ and express the spin correlation function as 
\begin{align}
G^R_0(\omega)=\frac{2S_0/\hbar}{\omega-\omega_{\bm{0}}+i\alpha_{\rm G} \omega} .
\end{align}

\subsection{Interfacial exchange interaction}
\label{sec:intmodel}

Now let us consider the interfacial exchange interaction between the FI and the CNT with the Hamiltonian,
\begin{align}
\mathcal{H}_{\rm int}=S_{\bm{0}}^+ s^- + S_{\bm{0}}^- s^+,
\end{align}
where $s^\pm$ is the spin ladder operator of the CNT, defined as
\begin{align}
s^- &= \sqrt{\frac{1}{N_{\rm FI}}} \sum_{r,r'} \sum_{\alpha,\alpha'} \int_0^W dx \, J(x) \nonumber \\
& \times e^{-i(\alpha-\alpha') k_{\rm F}x-i(r-r')q_{\rm F}x} \psi^\dagger_{r\alpha-}(x) \psi_{r'\alpha'+}(x), 
\label{tildeSp}
\end{align}
and $s^+ = (s^-)^{\dagger}$.
Here, $W$ is the length of the interface, $J(x)$ is the interfacial exchange coupling, $N_{\rm FI}$ is the number of unit cells in the FI, $k_{\rm F}$ is the Fermi wavenumber, and $q_{\rm F}$ ($\ll k_{\rm F}$) is the momentum mismatch associated with the two modes.
Because the interfacial exchange coupling $J(x)$, which is induced by quantum mechanical mixing between CNT and FI, is sensitive to distances of atoms across the junction, we assumed that it depends on the position $x$ due to random atomic configuration near the interface.
A simplified model for randomness in $J(x)$ will be accounted for in the next section.

\section{Formulation}
\label{sec:formuation}

\subsection{Gilbert damping}
\label{sec:Gilbert}

Using second-order perturbation with respect to the interfacial exchange coupling, the spin susceptibility is calculated as
\begin{align}
G(i\omega_n)&=\frac{1}{G_0(i\omega_n)^{-1}-\Sigma(i\omega_n)}\\
\Sigma(i\omega_n)&=-\frac{1}{\hbar}\int^{\hbar\beta}_0 d\tau e^{i\omega_n\tau}\langle T_\tau s^+(\tau) s^-(0) \rangle
\end{align}
where $s^\pm (\tau) = e^{\mathcal{H}_{\rm CNT}\tau/\hbar} s^\pm e^{-\mathcal{H}_{\rm CNT}\tau/\hbar}$.
The retarded spin correlation function is obtained by analytic continuation $i\omega_n \rightarrow \omega + i\delta$ as
\begin{align}
G^R(\omega)&=\frac{2S_0/\hbar}{\omega-(\omega_{\bm{0}}+\delta \omega_{\bm{0}}) + i(\alpha_{\rm G}+\delta \alpha_{\rm G})\omega_{\bm 0} }, \\
\frac{\delta \omega_{\bm 0}}{\omega_{\bm 0}} & \simeq \frac{2S_0}{\hbar \omega_{\bm 0}} \, {\rm Re} \, \Sigma^R(\omega_{\bm 0}), \\
\delta \alpha_{\rm G} &\simeq -\frac{2S_0}{\hbar \omega_{\bm 0}} \, {\rm Im} \, \Sigma^R(\omega_{\bm 0}), 
\end{align}
where $\Sigma^R(\omega)$ is the retarded self-energy defined by
\begin{align}
\Sigma^R(\omega) & = \int dt \, e^{i\omega t} \Sigma^R(t) , \\
\Sigma^R(t) &= -\frac{i\theta(t)}{\hbar} \langle [s^+(t),s^-(0)] \rangle,
\label{eq:SigmaRt}
\end{align}
$\theta(t)$ is the step function, and $\alpha_{{\rm G}} + \delta \alpha_{\rm G} \ll 1$ has been assumed.
In our work, we focus on the increase in the Gilbert damping due to the junction, $\delta \alpha_{\rm G}$, which is written in terms of the dynamic spin susceptibility of CNTs.

\subsection{Self-energy of electrons in CNTs}
\label{self energy}

By substituting Eq.~(\ref{tildeSp}) into Eq.~(\ref{eq:SigmaRt}), we obtain
\begin{align}
&\Sigma^R(t) = -\frac{i}{\hbar} \theta(t) \frac{2S_0}{N_{\rm FI}} \sum_{r,r'} \sum_{\alpha,\alpha'} \int_0^W dx \int_0^W dy \langle J(x)J(y) \rangle_{\rm imp} \nonumber \\
& \hspace{8mm} \times  e^{-i(k_{\rm F}(\alpha-\alpha') + q_{\rm F}(r-r'))(x-y)}C_{r\alpha r'\alpha'}(x,y,t) ,\\
&C_{r\alpha r'\alpha'}(x,y,t) = \langle [ \psi^\dagger_{r\alpha,+}(x,t) \psi_{r'\alpha',-}(x,t), \nonumber \\
& \hspace{30mm} \psi^\dagger_{r'\alpha',-}(y,0) \psi_{r\alpha,+}(y,0) ]\rangle_0 .
\label{eq:Cdef}
\end{align}
Here, $\langle \cdots\rangle_{\rm imp}$ indicates a random average for the interfacial exchange coupling.
For simplicity, we assume that the exchange coupling follows a Gaussian distribution whose average and variance are given by
\begin{align}
& \langle J(x) \rangle_{\rm imp} = J_1, \\
& \langle \delta J(x) \delta J(y) \rangle_{\rm imp} = J_2^2 a \delta(x-y) ,
\end{align}
where $\delta J(x) = J(x)-\langle J(x) \rangle_{\rm imp}$.
Here, $J_1$ and $J_2$ represent respectively the average and the standard deviation of the distribution. The ratio $J_2/J_1$ reflects the randomness of the interfacial exchange coupling.
In particular, the case of $J_2/J_1=0$ corresponds to a clean interface without randomness.

Accordingly, the self-energy is calculated as
\begin{align}
\Sigma^R(t) &= \Sigma_1^R(t) + \Sigma_2^R(t) , \\
\Sigma_1^R(t) &= -i\theta(t)
\frac{2S_0J_1^2}{\hbar N_{\rm FI}}\sum_{r,r',\alpha,\alpha'} \int_0^W dx \int_0^W dy \, \nonumber \\ &\times e^{-i(k_{\rm F}(\alpha-\alpha') + q_{\rm F}(r-r'))(x-y)} C_{r\alpha r'\alpha'}(x,y,t),\\
 \Sigma_2^R(t) &= -i\theta(t)
\frac{2S_0J_2^2a}{\hbar N_{\rm FI}} \sum_{r,r',\alpha,\alpha'} \int_0^W dx \,  C_{r\alpha r'\alpha'}(x,x,t). \label{Sigma2}
\end{align}
Since the integrand of $\Sigma_1^R(t)$ includes a rapidly oscillating part as a function of $(x-y)$, the integral is negligibly small except for the case of $\alpha=\alpha'$ and $r=r'$. There, we obtain
\begin{align}
\Sigma_1^R(t) &=-i\theta(t)
\frac{2S_0J_1^2}{\hbar N_{\rm FI}} \sum_{r,\alpha} \int_0^{W} dx \int_0^{W} dy \, C_{r\alpha r\alpha}(x,y,t) . 
\label{Sigma1}
\end{align}
We should note that $\Sigma_1^R(t)$ corresponds to the process of electron creation and annihilation in the same branch and represents momentum-conserving spin relaxation for a clean junction. 
In contrast, $\Sigma_2^R(t)$ represents spin relaxation for a ``dirty'' junction that is independent of the electron momentum.
Here, the word ``dirty'' means that during spin exchange process the momentum of electrons in the CNT is not conserved and transitions between different branches of valleys and propagation directions are allowed.
The following discussion will consider two limiting cases for the interface.
For the clean interface limit ($J_1 \gg J_2$), the magnon self-energy is represented with $\Sigma_1^R(t)$, while in the dirty interface limit ($J_1 \ll J_2$), it is represented with $\Sigma_2^R(t)$.

\subsection{Clean interface}

Since the correlation function $C_{r\alpha r' \alpha'}(x,y,t)$ can be calculated using the bosonization method (see Appendix~\ref{app:CorrelationFunctions}), the self-energy $\Sigma_1^R(t)$ can be obtained analytically.
Therefore, the corresponding increase in the Gilbert damping is obtained as
\begin{align}
& \delta \alpha_{{\rm G},1} = -\frac{2S_0}{\hbar \omega_{\bm 0}} {\rm Im} \, \Sigma_1^R(\omega_{\bm 0}) \nonumber \\
&= -\frac{4S_0J_1^2}{\hbar^2 \omega_{\bm 0}(2\pi a)^2N_{\rm FI}} \int_0^W \! \! dx \int_0^W \! \! dy \int_0^\infty \! \! dt \, \sin \omega_0 t \nonumber \\
&\times {\rm Im} \,
\Biggl[ \left( \frac{\sinh(i \pi a/\beta\hbar v_{\rm F})}{\sinh(\pi (ia-(x-y)-v_{\rm F}t)/\beta\hbar v_{\rm F})}\right)^{\gamma-1} \nonumber \\
& \hspace{2mm} \times \left( \frac{\sinh(i\pi a/\beta\hbar v_{\rm F})}{\sinh(\pi (ia+(x-y)-v_{\rm F}t)/\beta\hbar v_{\rm F})}\right)^{\gamma+1} \Biggr], \\
\gamma &\equiv \frac{K_{s+}}{4} + \frac{K_{s-}}{4}
+ \frac{1}{4K_{s+}}+ \frac{1}{4K_{s-}} .
\label{gammaexpression}
\end{align}
After analytic integration with respect to $t$  (see Appendix~\ref{app:integral} for details), we obtain
\begin{align}
\delta \alpha_{{\rm G},1} &= \frac{2}{\pi}
\frac{\Gamma(\gamma)^2}{\Gamma(2\gamma)}
\frac{S_0 J_1^2 W a}{\hbar^2 v_{\rm F}^2 N_{\rm FI}} \left(\frac{2\pi a}{\beta \hbar v_{\rm F}}\right)^{2\gamma-3} \nonumber \\
&\hspace{7mm} \times I(\pi W/\beta \hbar v_{\rm F},\gamma), 
\label{cleanana1}
\\
I(w,\gamma) &= \frac{1}{w} \int^w_0 dz' \int^{z'}_0 dz \, e^{-2(\gamma-1)z} \nonumber \\
&\hspace{7mm} \times F(\gamma-1,\gamma,2\gamma; 1-e^{-4z}) ,
\label{cleanana2}
\end{align}
where $F(a,b,c;x)$ is the hypergeometric function.

\subsection{Dirty interface}

The self-energy $\Sigma_2^R(t)$ can be obtained in a similar way as above.
The corresponding increase in the Gilbert damping is given by
\begin{align}
& \delta \alpha_{{\rm G},2} = -\frac{2S_0}{\hbar \omega_{\bm 0}} {\rm Im} \, \Sigma_2^R(\omega_{\bm 0}) \nonumber \\
&= -\frac{S_0 J_2^2 a W}{\hbar^2 \omega_{\bm 0} (\pi a)^2 N_{\rm FI}} \sum_{r,r',\alpha,\alpha'} \int_0^\infty dt \, \sin \omega_{\bm 0}t \nonumber \\
& \hspace{2mm} \times {\rm Im} \, \left[ \left(\frac{\sinh(i\pi a/\beta\hbar v_{\rm F})}{\sinh(\pi(ia-v_{\rm F}t)/\beta\hbar v_{\rm F})} \right)^{2\gamma_{r\alpha r'\alpha'}} \right],
\label{eq:AnalyticFormDirty} \\
& \gamma_{r\alpha r'\alpha'} = \gamma_1 \delta_{r,r'} \delta_{\alpha,\alpha'} + \gamma_2 \delta_{r,-r'} \delta_{\alpha,\alpha'} \nonumber \\
& \hspace{20mm} + \gamma_3 \delta_{r,r'} \delta_{\alpha,-\alpha'} + \gamma_4 \delta_{r,-r'} \delta_{\alpha,-\alpha'} , 
\label{eq:gammara} \\
& \gamma_1 = (K_{s+} + K_{s-} + 1/K_{s+} + 1/K_{s-})/4, \\
& \gamma_2 = (K_{c+} + K_{c-} + 1/K_{s+} + 1/K_{s-})/4, \\
& \gamma_3 = (K_{s+} + K_{c-} + 1/K_{c+} + 1/K_{s-})/4, \\
& \gamma_4 = (K_{c+} + K_{s-} + 1/K_{s+} + 1/K_{c-})/4. 
\end{align}
We should note that $\delta \alpha_{{\rm G},2}$ is proportional to $W$, since the spin relaxation rate is determined through spatially-local spin exchange in the dirty interface and is proportional to the number of spin-exchange channels.
After analytic integration with respect to $t$ (see Appendix~\ref{app:integral} for details), we obtain
\begin{align}
\delta \alpha_{{\rm G},2} &= \frac{1}{2\pi}
\frac{S_0J_2^2 a W}{\hbar^2 v_{\rm F}^2 N_{\rm FI}} 
\nonumber \\
&\times \sum_{r,r',\alpha,\alpha'} \frac{\Gamma(\gamma_{r\alpha r'\alpha'})^2}{
\Gamma(2\gamma_{r\alpha r'\alpha'})} \left(\frac{2\pi a}{\beta \hbar v_{\rm F}}\right)^{2\gamma_{r\alpha r'\alpha'}-2} .
\end{align}

\section{Numerical Estimate}
\label{sec:result}

Next, we evaluate numerically the increase in the Gilbert damping by using realistic experimental parameters.
While the increase was formulated for a single CNT in the previous section, to increase the signal, it would be more useful if we considered a junction with a bundle of CNTs.
Thus, in the following, we will consider a junction composed of a FI and a bundle of CNTs with an area of $W\times W'$ (see Fig.~\ref{fig:spin-pumping-setup}) and multiply $\delta \alpha_{{\rm G},1}$ and $\delta \alpha_{{\rm G},2}$ by the number of CNTs in the junction, $N_{\rm CNT}=W'/d$ ($d$: the diameter of CNTs).

\begin{table}[tb]
 \caption{Parameters used for the numerical estimate.}
 \label{table:Parameters}
 \centering
  \begin{tabular}{lll}
   \hline
   Microwave frequency & $\omega_{\bm 0}$ & $1\, {\rm GHz}$ \\
   Fermi velocity of CNT & $v_{\rm F}$ & $10^6\, {\rm m}/{\rm s}$ \\
   Lattice constant of CNT & $a$ & $2.46\,$\AA \\
   Diameter of CNT & $d$ & $1.5 \, {\rm nm}$ \\
   Amplitude of spins of FI & $S_0$ & 10 \\
   Lattice constant of FI & $a'$ & $12.376\,$\AA \\
   Thickness of FI & $d'$ & $10\, {\rm nm}$ \\
   Interfacial exchange couplings & $J_1$, $J_2$ & \\
   \ \ \ clean interface &  & $J_1=2\, {\rm K}$, $J_2=0$ \\
   \ \ \ dirty interface & & $J_1=0\, {\rm K}$, $J_2=1,2,3 \, {\rm K}$ \\
   Luttinger parameters & $K_{c+}$ & 0.20 \\
    & $K_{s+}$ & 1.07 \\
    & $K_{c-}$, $K_{s-}$ & 1 \\
   \hline
  \end{tabular}
\end{table}

The parameters are given in Table~\ref{table:Parameters}.
The Fermi velocity $v_{\rm F}$, lattice constant $a$, diameter $d$, Luttinger parameters of CNTs, $K_{c+}$, $K_{c-}$, and $K_{s-}$, are taken from Refs.~\cite{Giamarchi2003,Egger1998,Yoshioka1999}.
The value of $K_{s+}$ is an experiment result~\cite{Dora2007} under a magnetic field of $3.6 \, {\rm T}$\footnote{We treated effect of the external magnetic field and spin-orbit interaction phenomenologically through a change in $K_{s+}$.}.
The spin amplitude $S_0$ and the lattice constant $a'$ are determined by assuming that the FI is made from yttrium iron garnet (YIG). 
The interfacial exchange coupling ($J_1$ or $J_2$) is roughly estimated to be $2 \, {\rm K}$~\cite{Nogues1999}. 
The number of unit cells is estimated as $N_{\rm FI} = WW'd'/a'^3$, where $d'$ is the thickness of the FI.

\subsection{Clean interface}

\begin{figure}[tbp]
  \begin{center}
    \includegraphics[clip, width=8.5cm]{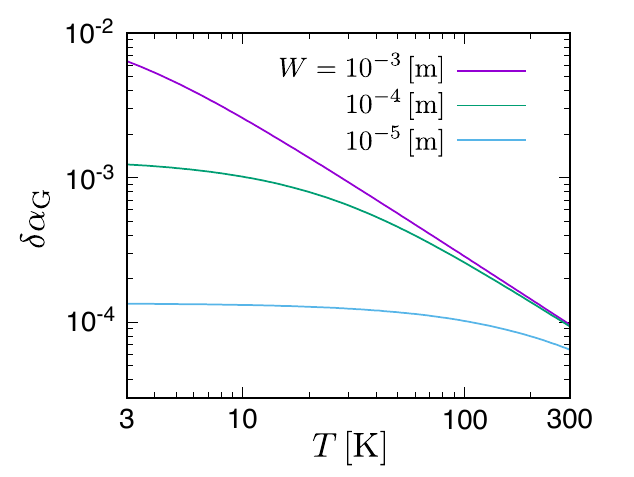}
    \caption{Temperature dependence of increase in the Gilbert damping, $\delta \alpha_{{\rm G},1}$, for a clean interface ($J_1 \gg J_2$).}
    \label{fig:CleanT}
  \end{center}
\end{figure}

\begin{figure}[tbp]
  \begin{center}
    \includegraphics[clip, width=8.5cm]{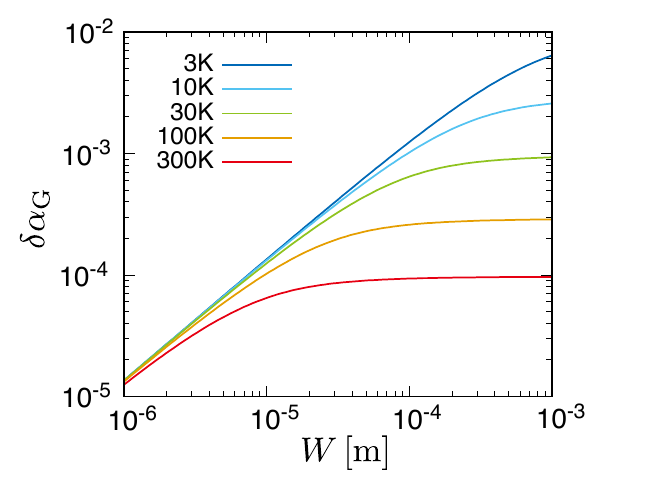}
    \caption{Junction-length dependence of increase in the Gilbert damping, $\delta \alpha_{{\rm G},1}$, for a clean interface ($J_1 \gg J_2$).}
    \label{fig:CleanW}
  \end{center}
\end{figure}

The estimated increase in the Gilbert damping for a clean interface ($J_1 = 2 \, {\rm K} \gg J_2$) is shown in Fig.~\ref{fig:CleanT} as a function of temperature.
While $\delta\alpha_{G,1}$ is proportional to $1/T$ at high temperatures, it is almost constant at low temperatures.
The crossover temperature for a fixed length $W$ is given by $T^* = g(\gamma) \hbar v_{\rm F}/(k_{\rm B}W)$ ($k_{\rm B}$: Boltzmann constant), which is proportional to $1/W$.
The factor $g(\gamma)$, which depends only on $\gamma$, is explicitly shown later.  
The increase in the Gilbert damping is shown as a function of the junction length $W$ in Fig.~\ref{fig:CleanW}.
While $\delta\alpha_{G,1}$ is proportional to $W$ for a short junction, it is almost constant for a long junction.
The crossover length for a fixed temperature $T$ is given by $W^*=g(\gamma) \hbar v_{\rm F}/( k_{\rm B} T)$.

In the present estimate, the condition  $L_{\rm th} \ll v_{\rm F}/\omega_{\bm 0}$ always holds, where $L_{\rm th}=\hbar v_{\rm F}/k_{\rm B}T$ is a thermal length.
Under this condition, the increase in the Gilbert damping becomes independent of $\omega_{\bm 0}$ and is approximately given by
\begin{align}
& \delta \alpha_{G,1} = \frac{\Gamma(\gamma)^2}{\Gamma(2\gamma)} \frac{S_0J_1^2 a'^3 a}{(\hbar v_{\rm F})^{2}dd'} \left(\frac{2\pi a}{{L_{\rm th}}}\right)^{2\gamma-3} \!\!\! f(\gamma,\pi W/L_{\rm th}) , \label{eq:approx1} \\ 
&f(\gamma,w) = \left\{ \begin{array}{ll}
w/\pi, & \displaystyle{
(w/\pi \ll  g(\gamma))} ,\\
g(\gamma), & 
(w/\pi \gg g(\gamma)), 
\end{array}
\right.\label{eq:approx2} \\
&g(\gamma) = \frac{2}{\pi} \int_0^\infty dz \,
e^{-2(\gamma-1)z} F(\gamma-1,\gamma,2\gamma; 1-e^{-4z}).
\label{eq:approx3}
\end{align}
From this analytic expression, we obtain
\begin{align}
\delta \alpha_{G,1} &\propto \left\{ \begin{array}{ll}
T^{2\gamma-2} W, & (W \ll g(\gamma)L_{\rm th}), \\
T^{2\gamma-3} g(\gamma), & \displaystyle{
(W \gg g(\gamma)}L_{\rm th}).
\end{array}
\right. 
\end{align}
The exponent $\gamma =(K_{s+}+K_{s-}+K_{s+}^{-1}+K_{s-}^{-1})/4$ corresponds to unity when $K_{s+}=K_{s-}=1$.
Even in the present estimate employing $K_{s+} = 1.07$, the exponent is almost unity ($\gamma = 1.00114$).
By setting $\gamma = 1$, we can reproduce the power in the temperature and junction-length dependence of $\delta \alpha_{G,1}$ shown in Figs.~\ref{fig:CleanT} and \ref{fig:CleanW}.

Finally, let us discuss the factor $g(\gamma)$. If $\gamma$ is slightly larger than 1 as in the present estimate, the geometric function is approximated as $F(\gamma-1,\gamma,2\gamma;x) \simeq 1$. Then, the factor $g(\gamma)$ is approximately given as
\begin{align}
g(\gamma) = \frac{1}{\pi(\gamma-1)}.
\end{align}
This expression indicates that the increase in the Gilbert damping in the high-temperature limit ($T \gg T^*$) or the long-junction limit ($W \gg W^*$) is highly sensitive to the deviation of $\gamma$ from unity.
The crossover temperature $T^*$ and the crossover length $W^*$ also include the factor $g(\gamma) \propto (\gamma-1)^{-1}$.
Thus, the increase in the Gilbert damping can be used
to investigate small deviations of $\gamma$ from unity. 
Then, the Luttinger parameter $K_{s,+}$ in the spin sector can also be determined from Eq.~(\ref{gammaexpression}) if we know whether it is greater or less than unity.
We note that in the NMR measurement~\cite{Dora2007} $K_{s,+}$ decreases as the magnetic field increases.
Using this experimental tendency, we expect that $K_{s,+}$ can be determined uniquely.

\subsection{Dirty interface}

\begin{figure}[tbp]
  \begin{center}
    \includegraphics[clip, width=8cm]{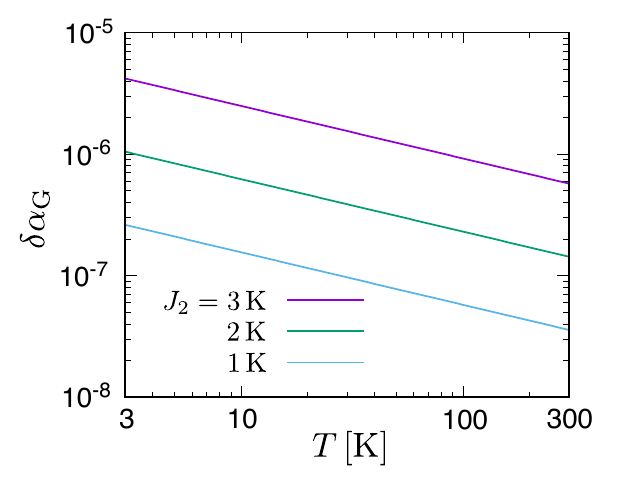}
    \caption{Temperature dependence of increase in the Gilbert damping, $\delta \alpha_{{\rm G},2}$, for a dirty interface ($J_2 \gg J_1$).
    The three lines correspond to $J_2 = 1$, 2, and $3 \, {\rm K}$, respectively.}
    \label{fig:DirtyT}
  \end{center}
\end{figure}

Next, we consider a dirty interface ($J_2 \gg J_1$).
Figure~\ref{fig:DirtyT} shows the increase of the Gilbert damping, $\delta \alpha_{{\rm G},2}$, as a function of the temperature for $J_2 = 1$, $2$, and $3 \, {\rm K}$.
In this case, $\delta\alpha_{G,2}$ is proportional to $T^{-0.43}$ in the whole temperature range and shows a nontrivial exponent inherent to the Tomonaga-Luttinger liquid.

The condition $L_{\rm th} \ll v_{\rm F}/\omega_{\bm 0}$ also holds for a dirty interface. 
Therefore, $\delta \alpha_{{\rm G},2}$ can be approximated as
\begin{align}
\delta \alpha_{G,2} &= \frac{1}{2\pi} 
\frac{S_0J_2^2a a'^3}{(\hbar v_{\rm F})^2 dd'} \nonumber \\
&\times \sum_{r,r',\alpha,\alpha'} \frac{\Gamma(\gamma_{r\alpha r'\alpha'})^2}{\Gamma(2\gamma_{r\alpha r'\alpha'})}\left(\frac{2\pi a}{L_{\rm th}}\right)^{2\gamma_{r\alpha r'\alpha'}-2} .
\label{eq:ag2approx}
\end{align}
Noting that $a\ll L_{\rm th}$, the factor $(2\pi a/L_{\rm th})^{2\gamma_{r\alpha r'\alpha'}}$ in Eq.~(\ref{eq:ag2approx}) is largely reduced as $\gamma_{r\alpha r'\alpha'}$ increases.
Therefore, in the sum of Eq.~(\ref{eq:ag2approx}), it is sufficient to keep the terms in which $\gamma_{r\alpha r'\alpha'}$ takes a minimum value.
In the present estimate, $\gamma_{r\alpha r'\alpha'}$ is given by Eq.~(\ref{eq:gammara}) with
\begin{align}
(\gamma_1,\gamma_2,\gamma_3,\gamma_4) &= (1.001,0.784,1.001.0.784). 
\end{align}
Upon setting the minimum exponent to be $\gamma_{\rm min} = 0.784$, we obtain $\delta \alpha_{G,2} \propto T^{2\gamma_{\rm min}-2} = T^{-0.432}$, which is consistent with the numerical results shown in Fig.~\ref{fig:DirtyT}.
Therefore, the nontrivial exponent inherent to the Tomonaga-Luttinger liquid appears in spin pumping through a dirty junction. 
Note that the approximate expression is independent of the junction length $W$ for a fixed thickness, since the $W$-linear factor in $N_{\rm FI}=WW'd/a'^3$ cancels out the factor of $W$ in Eq.~(\ref{eq:AnalyticFormDirty}).

The equation for the increase in the Gilbert damping for the dirty interface has almost the same form as that for $1/T_1T$ in NMR experiments where $T_1$ is the longitudinal relaxation time of nuclear spins\cite{Singer2005,Dora2007,Ihara2010}. 
Therefore, the power law of the temperature dependence for the dirty interface is the same as in NMR experiments.
This is because the spin transfer occurs at a spatially localized point due to the impurity average at the dirty interface, leading to the same situation as the NMR experiment in which $1/T_1T$ is related to the local dynamic spin susceptibility.

\section{Experimental Relevance}
\label{sec:Relevance}

We estimated the increase in the Gilbert damping $\delta \alpha_{\rm G}$ in two limiting situations, i.e., clean and dirty interfaces. 
If we choose YIG as the ferromagnet, $\delta \alpha_{\rm G}$ should be roughly in the range $10^{-5}$--$10^{-2}$, because it should be comparable to the Gilbert damping of bulk YIG, $\alpha_{\rm G}$, which is of order of $10^{-5}$--$10^{-3}$.
For a clean interface, $\delta \alpha_{\rm G}$ is large enough to be measured in FMR experiments (see Figs.~\ref{fig:CleanT} and \ref{fig:CleanW}).
Note that $\delta \alpha_{\rm G}$ can be reduced by increasing the thickness of YIG (denoted by $d'$).
On the other hand, for a dirty interface, $\delta \alpha_{\rm G}$ is too small for it to be observable by spin pumping (see Fig.~\ref{fig:DirtyT}).
However, we will moderate judgement on the possibility of observing $\delta \alpha_{\rm G}$ for a dirty interface, because detailed information on the interfacial exchange coupling is still lacking.
We should note that in the present modeling of randomness, the increase in the Gilbert damping is given by a sum of these two contributions, i.e., $\delta \alpha_{\rm G}=\delta \alpha_{{\rm G},1}+\delta \alpha_{{\rm G},2}$, for an arbitrary strength of interfacial randomness.

Our calculation can be applied straightforwardly to other one-dimensional electron systems such as quasi-one-dimensional magnets, whose low-energy states are also described by the Tomonaga-Luttinger liquid model.
In particular, the low-energy states of spin systems with in-plane anisotropy are characterized by a Luttinger parameter $K_s$ smaller than 1.
If $K_s$ is sufficiently smaller than 1, $\delta \alpha_{\rm G}$ should show nontrivial power-law behavior with respect to the temperature even for a clean interface.

\section{Summary}
\label{sec:summary}

We theoretically studied spin pumping from a ferromagnetic insulator into carbon nanotubes.
First, we formulated the increase in the Gilbert damping in terms of the spin susceptibility and described the interfacial exchange coupling with a simple model, in which two types of spin-flip process, i.e., momentum-conserving and momentum-nonconserving processes, coexist.
Then, we analytically calculated the increase in the Gilbert damping by treating electrons in carbon nanotubes in the framework of the Luttinger liquid.
For a clean interface, the increase in damping is proportional to the inverse of the temperature at high temperatures while it is almost constant at low temperatures.
The crossover temperature includes information on the Fermi velocity in carbon nanotubes.
We also found that the increase in damping is highly sensitive to the deviation of the Luttinger parameter in the spin sector from unity.
For a dirty interface, the increase in damping shows a power-law dependence on the temperature with a nontrivial exponent reflecting the nature of the Tomonaga-Luttinger liquid.
We also estimated the increase of the Gilbert damping using realistic parameters.
Our results indicate a possible application of spin pumping for detecting power-law behavior of spin excitation in low-dimensional systems.
Detection of other types of spin excitation in exotic many-body states will be left as a future study.

\acknowledgments

This French-Japanese collaboration is supported by the CNRS International Research Project ``Excitations in Correlated Electron Systems driven in the GigaHertz range'' (ESEC).
This work received support from the French government under the France 2030 investment plan, as part of the Initiative d'Excellence d'Aix-Marseille Universit\'{e} - A*MIDEX.
We acknowledge support from the institutes IPhU (AMX-19-IET-008) and AMUtech (AMX-19-IET-01X).
T. K. acknowledges support from the Japan Society for the Promotion of Science (JSPS KAKENHI Grants No. 20K03831).
M. M. acknowledges support by a Grant-in-Aid for Scientific Research B (23H01839 and 21H01800) and A (21H04565) from MEXT, Japan, and by the Priority Program of Chinese Academy of Sciences, under Grant No. XDB28000000.

\appendix

\section{Correlation functions}
\label{app:CorrelationFunctions}

Here, we briefly summarize the calculation of the correlation function $C_{r\alpha r'\alpha'}(x,y,t)$ defined in Eq.~(\ref{eq:Cdef}).
Using the bosonic fields, the correlation function is written as
\begin{align}
& C_{r\alpha r'\alpha'}(x,y,t) \nonumber \\
&=  \frac{1}{(2\pi a)^2} \left[ \langle
e^A e^B e^C e^D\rangle_0 - \langle
e^C e^D e^A e^B\rangle_0 \right] ,\\
&A = -i(-r\theta_{\alpha +}(x,t)+\phi_{\alpha +}(x,t)) , \\
&B = i(-r'\theta_{\alpha'-}(x,t)+\phi_{\alpha'-}(x,t)) ,\\
&C = -i(-r'\theta_{\alpha'-}(y,0)+\phi_{\alpha'-}(y,0)) ,\\
&D = i(-r\theta_{\alpha +}(y,0)+\phi_{\alpha +}(y,0)), \end{align}
where we set $r=+1$ ($r=-1$) for the left-going (right-going) branch.
Using the formula,
\begin{align}
\langle e^{A_1} e^{A_2} \cdots e^{A_N} \rangle &=  \exp \left[ \frac12 \sum_{i} \langle A_i^2\rangle  + \sum_{i<j} \langle A_i A_j \rangle \right] ,
\end{align}
which holds when $[A_i,A_j]$ is a $c$-number, we obtain
\begin{align}
&\langle e^A e^B e^C e^D\rangle_0 \equiv e^{F_{r\alpha r' \alpha'}(x-y,t)} \nonumber \\
&= \exp \Bigl[ \frac12 \langle (A^2+B^2+C^2+D^2) \rangle + \langle A B \rangle + \langle C D \rangle \nonumber \\
& \hspace{12mm} + \langle A C \rangle + 
\langle A D \rangle 
+ \langle B C \rangle + \langle B D \rangle  
\Bigr] , \\
&\langle e^C e^D e^A e^B\rangle_0 = e^{F_{r'\alpha' r \alpha}(y-x,-t)} .
\end{align}
The correlation functions of the bosonic fields, which are defined as $G_{j\delta}^{XY}=\langle X(x,t) Y(y,0)\rangle$, are calculated as~\cite{vonDelft1998}
\begin{align}
G_{j\delta}^{\theta\theta}(x,t) &= \frac{K_{j\delta}}{4} (I(x,t) + I(-x,t)), \\
G_{j\delta}^{\phi\phi}(x,t) &= \frac{1}{4K_{j\delta}}  (I(x,t) + I(-x,t)), \\
G_{j\delta}^{\theta\phi}(x,t) &= G_{j\delta}^{\phi\theta}(x,t) = \frac14 (I(x,t) - I(-x,t)), \\
I(x,t) &= -\log\left[ \frac{2i\beta \hbar v_{\rm F}}{L} \sinh \left( \frac{\pi (ia -x-v_{\rm F}t)}{\beta \hbar v_{\rm F}} \right) \right] .
\end{align}
Using these correlation functions, we obtain
\begin{align}
F_{r\alpha r' \alpha'}(x,t) &= F_1 \delta_{r,r'} \delta_{\alpha,\alpha'} + F_2 \delta_{r,-r'} \delta_{\alpha,\alpha'} \nonumber \\
&\hspace{5mm} + F_3 \delta_{r,r'} \delta_{\alpha,-\alpha'}
+ F_4 \delta_{r,-r'} \delta_{\alpha,-\alpha'} , \\
F_1(x,t) &= \tilde{G}^{\theta\theta}_{s+}+\tilde{G}^{\theta\theta}_{s-}+\tilde{G}^{\phi\phi}_{s+}+\tilde{G}^{\phi\phi}_{s-} \nonumber \\
&\hspace{5mm} -r(\tilde{G}^{\theta\phi}_{s+}+\tilde{G}^{\theta\phi}_{s-}+\tilde{G}^{\phi\theta}_{s+}+\tilde{G}^{\phi\theta}_{s-}) , \\
F_2(x,t)  &= \tilde{G}^{\theta\theta}_{c+}+\tilde{G}^{\theta\theta}_{c-}+\tilde{G}^{\phi\phi}_{s+}+\tilde{G}^{\phi\phi}_{s-} , \\
F_3(x,t)  &= \tilde{G}^{\theta\theta}_{s+}+\tilde{G}^{\theta\theta}_{c-}+\tilde{G}^{\phi\phi}_{s+}+\tilde{G}^{\phi\phi}_{c-}\nonumber \\
&\hspace{5mm} -r(\tilde{G}^{\theta\phi}_{s+}+\tilde{G}^{\theta\phi}_{c-}+\tilde{G}^{\phi\theta}_{s+}+\tilde{G}^{\phi\theta}_{c-}) , \\
F_4(x,t)  &= \tilde{G}^{\theta\theta}_{c+}+\tilde{G}^{\theta\theta}_{s-}+\tilde{G}^{\phi\phi}_{s+}+\tilde{G}^{\phi\phi}_{c-},
\end{align}
where $\tilde{G}^{XY}_{j\delta}(x,t) \equiv G^{XY}_{j\delta}(x,t)-G^{XY}_{j\delta}(0,0)$.
Combining these results enables the correlation function $C_{r\alpha r'\alpha'}(x,y,t)$ to be obtained analytically.

\section{Analytic expressions of integrals}
\label{app:integral}

For a clean interface, the increase in damping is given as
\begin{align}
\delta \alpha_{{\rm G},1} &= - \frac{4S_0J_1^2}{\hbar^2 \omega_{\bm 0}(2\pi a)^2 N_{\rm FI}} {\cal I}_{\gamma}, 
\label{cleanapp1} \\
{\cal I}_{\gamma} &= v_{\rm F}^2 \left(\frac{\beta \hbar}{\pi} \right)^3 \int_0^w dx' \int_0^w dy' \int_0^\infty du \, \sin(\tilde{\omega}_0 u) \nonumber \\
&\times {\rm Im} \Biggl\{
\left[\frac{\sinh(i\alpha)}{\sinh(i\alpha+x-y-u)} \right]^{\gamma+1} \nonumber \\
&\hspace{7mm} \times 
\left[\frac{\sinh(i\alpha)}{\sinh(i\alpha-x+y-u)} \right]^{\gamma-1}
\Biggr\} ,
\end{align}
where $\tilde{\omega}_{\bm 0} = \beta \hbar \omega_{\bm 0}/\pi$, $u=\pi t/\beta \hbar$, $w=\pi W/\beta \hbar v_{\rm F}$, $x'=\pi x/\beta \hbar v_{\rm F}$, $y'=\pi y/\beta \hbar v_{\rm F}$, and $\alpha = \pi a/\beta \hbar v_{\rm F}$.
Changing variables from $x'$ and $y'$ with $Z=(x+y)/2$ and $z=x-y$, the integral is modified as
\begin{align}
{\cal I}_{\gamma} & =
v_{\rm F}^2 \left(\frac{\beta\hbar}{\pi}\right)^3
\int_0^\infty du \,  \sin(\tilde{\omega}_0 u) \nonumber \\ 
&\times \left[ \int_0^{w/2} dZ \int_{-2Z}^{2Z} dz + 
\int_{w/2}^w dZ \int_{-2(w-Z)}^{2(w-Z)} dz \right] \nonumber \\
&\times {\rm Im} \Biggl\{
\left[\frac{\sinh(i\alpha)}{\sinh(i\alpha+z-u)} \right]^{\gamma+1} \nonumber \\
&\hspace{7mm} \times 
\left[\frac{\sinh(i\alpha)}{\sinh(i\alpha-z-u)} \right]^{\gamma-1}
\Biggr\} \nonumber \\
&= - \frac{v_{\rm F}^2}{4} \left(\frac{\beta\hbar}{\pi}\right)^3 \int_0^{w/2} dZ \int_{2z}^{-2z} dz \int_{-\infty}^\infty du \nonumber \\
& \times ( e^{i\tilde{\omega}_0 u} - 
 e^{-i\tilde{\omega}_0 u} ) \nonumber \\
& \times \left[ \frac{\sinh(i\alpha)}{\sinh(i\alpha + z - u)}\right]^{\gamma + 1}
\left[ \frac{\sinh(i\alpha)}{\sinh(i\alpha - z - u)}\right]^{\gamma - 1}.
\end{align}
In the last equation, we have  used the relation
\begin{align}
\left( \frac{\sinh(i\alpha)}{\sinh(i\alpha\pm z-u)} \right)^*
= \frac{\sinh(i\alpha)}{\sinh(i\alpha\mp z+u)} 
\end{align}
and the symmetry of the integrand with respect to $Z \leftrightarrow w-Z$ and $z\leftrightarrow -z$.
At this stage, it is useful to introduce
\begin{align}
{\cal B}_\gamma(\zeta,z) &=
\int_{-\infty}^\infty du \, e^{i\zeta u} \left[ \frac{\sinh (i\alpha)}{\sinh(i\alpha+z-u)} \right]^{\gamma+1} \nonumber \\
&\times \left[ \frac{\sinh (i\alpha)}{\sinh(i\alpha-z-u)} \right]^{\gamma-1},
\end{align}
so that
\begin{align}
{\cal I}_\gamma &= -\frac{v_{\rm F}^2}{2} \left(\frac{\beta \hbar}{\pi}\right)^3
\int_0^{w/2} dZ \int_{-2Z}^{2Z} dz \nonumber \\
& \times [{\cal B}_\gamma(\tilde{\omega}_{\bm 0},z)-{\cal B}_\gamma(-\tilde{\omega}_{\bm 0},z)].
\end{align}
Setting $v=2u$ and rearranging the hyperbolic sine function, we obtain
\begin{align}
&{\cal B}_{\gamma}(\zeta,z) = \frac{1}{2} (1-e^{-2i\alpha})^{2\gamma} e^{-2z} \int_{-\infty}^\infty dv \nonumber \\
&\times 
\frac{e^{-v(\gamma-i\zeta/2)}}{[e^{-v}+e^{-2z+i(\pi-2\alpha)}]^{\gamma+1}[e^{-v}+e^{-2z+i(\pi-2\alpha)}]^{\gamma-1}},
\end{align}
which can be computed analytically, invoking the formula 3.315.1 in Ref.~\cite{Gradshteyn2007} as
\begin{align}
& {\cal B}_\gamma(\zeta,z) \nonumber \\
&= \frac{1}{2} (1-e^{-2i\alpha})^{2\gamma}e^{-i(\pi-2\alpha)(\gamma+i\zeta/2)} e^{-2z(\gamma-1-i\zeta/2)} \nonumber \\
&\times \frac{|\Gamma(\gamma+i\zeta/2)|^2}{\Gamma(2\gamma)} F(\gamma-1,\gamma,2\gamma;1-e^{-4z}),
\end{align}
where $F(a,b,c;x)$ is the Gauss hypergeometric function.
To leading order in the small parameter $\alpha$ ($\ll 1$), this reduces to
\begin{align}
{\cal B}_\gamma(\zeta,z)&= \frac{1}{2} (2\alpha)^{2\gamma}e^{\pi\zeta/2+i\zeta z} e^{-2z(\gamma-1)} \nonumber \\
&\hspace{-5mm} \times \frac{|\Gamma(\gamma+i\zeta/2)|^2}{\Gamma(2\gamma)} F(\gamma-1,\gamma,2\gamma;1-e^{-4z}).
\end{align}
In practice, we are mostly interested in the regime where $\tilde{\omega}_{\bm 0}=\beta \hbar \omega_{\bm 0} \ll 1$ so we can focus on values of $\zeta$ such that $|\zeta|\ll 1$. 
This allows us to expand $B_\gamma(\zeta,z)$ for small values of $\zeta$, which yields, after substituting back into the expression for ${\cal I}_\gamma$,
\begin{align}
{\cal I}_{\gamma} &= - \frac{\pi}{2} v_{\rm F}^2 \omega_{\bm 0} \left(\frac{\beta\hbar}{\pi}\right)^4 (2\alpha)^{2\gamma}
\frac{\Gamma(\gamma)^2}{\Gamma(2\gamma)} \int_0^{w/2} dZ
\nonumber \\
& \times \int_0^{2Z} dz e^{-2z(\gamma-1)} F(\gamma-1,\gamma,2\gamma; 1-e^{-4z}),
\end{align}
where we have used the symmetry of the integrand with respect to $z\leftrightarrow -z$.
Combining this result with Eq.~(\ref{cleanapp1}), Eqs.~(\ref{cleanana1}) and (\ref{cleanana2}) can be derived, after rewriting the integral variable as $z'=2Z$.

In the limiting case, the integral $I(w,\gamma)$ given in Eq.~(\ref{cleanana2}) can be approximated into a simple form.
For the short-junction limit ($w/\pi = W/\beta \hbar v_{\rm F} = W/L_{\rm th} \ll g(\gamma)$), we obtain
\begin{align}
I(w,\gamma) \simeq \frac{1}{w} \int_0^{w} dz' \int_0^{z'} dz ^, e^{-2z(\gamma-1)} = \frac{w}{2}.
\end{align}
For the long-junction limit ($w/\pi\gg g(\gamma)$),
\begin{align}
&I(w,\gamma)  \nonumber \\
& \simeq \frac{1}{w}
 \int_0^{w} dz' \int_0^{\infty} dz ^, e^{-2z(\gamma-1)} 
F(\gamma-1,\gamma,2\gamma; 1-e^{-4z}) \nonumber \\
&= \frac{\pi}{2} g(\gamma),
\end{align}
where $g(\gamma)$ is defined by Eq.~(\ref{eq:approx3}).
These analytical expressions lead to Eqs.~(\ref{eq:approx1}) and (\ref{eq:approx2}) in the main text.

For a dirty interface,
the increase in damping is expressed as
\begin{align}
&\delta \alpha_{{\rm G},2} = - \frac{S_0J_2^2aW}{\hbar^2 \omega_{\bm 0}(\pi a)^2 N_{\rm FI}} \sum_{r,r',\alpha,\alpha'}
{\cal I}_{\gamma_{r\alpha r'\alpha'}}' , \\
&{\cal I}_\gamma' = \frac{\beta \hbar}{\pi} \int_0^\infty du \, \sin(\tilde{\omega}_{\bm 0} u)
{\rm Im} \Biggl\{
\left[ \frac{\sinh(i\alpha)}{\sinh(i\alpha-u)} \right]^{2\gamma}
\Biggr\} ,
\end{align}
with the same dimensionless variables as for a clean interface.
By a similar way as the clean case, the integral ${\cal I}_{\gamma}'$ is modified as
\begin{align}
{\cal I}_{\gamma}' &= - \frac{\beta \hbar}{4\pi}
[A_{\gamma}(-\tilde{\omega}_{\bm 0})-A_{\gamma}(-\tilde{\omega}_{\bm 0})], \\
A_\gamma(\zeta) &= \int_{-\infty}^\infty du \, e^{-i\zeta u} \left[ \frac{\sinh(i\alpha)}{\sinh(i\alpha-u)}\right]^{2\gamma}.
\end{align}
Setting $v=2u$ and rearranging the hyperbolic sine function, we obtain
\begin{align}
A_{\gamma}(\zeta) &= \frac{1}{2}(1-e^{-2i\alpha})^{2\gamma} \int_{-\infty}^\infty dv \frac{e^{-(\gamma+i\zeta/2)v}}{(e^{-v}+e^{i(\pi-2\alpha)})^{2\gamma}}.
\end{align}
Invoking the formula 3.314 in Ref.~\cite{Gradshteyn2007}, this can be computed as
\begin{align}
A_\gamma(\zeta)=\frac{1}{2} (2\sin \alpha)^{2\gamma} e^{\alpha \zeta} e^{-\pi\zeta/2}
\frac{|\Gamma(\gamma+i\zeta/2)|^2}{\Gamma(2\gamma)},
\end{align}
which then yields, to leading order in $\alpha$ ($\ll 1$), 
\begin{align}
{\cal I}_{\gamma}' = -\frac{\beta\hbar}{4\pi} (2\alpha)^{2\gamma}
\frac{|\Gamma(\gamma+i\zeta/2)|^2}{\Gamma(2\gamma)}
\sinh(\pi \tilde{\omega}_{\bm 0}/2).
\end{align}
Assuming $\tilde{\omega}_{\bm 0}=\beta \hbar \omega_{\bm 0}/\pi \ll 1$, this is further simplified as
\begin{align}
{\cal I}_{\gamma}' = - \frac{\beta^2\hbar^2}{8\pi} \omega_{\bm 0} \left(\frac{2\pi a}{\beta \hbar v_{\rm F}}\right)^{2\gamma} 
\frac{\Gamma(\gamma)^2}{\Gamma(2\gamma)},
\end{align}
which finally leads to Eq.~(\ref{eq:ag2approx}) in the main text.

\bibliography{ref}

\end{document}